\documentclass[reprint,prl,amsmath,amssymb,superscriptaddress,floatfix]{revtex4-2}
\usepackage[dvipdfmx]{graphicx}
\usepackage[dvipsnames]{xcolor}
\usepackage[colorlinks,linkcolor=Blue,citecolor=Blue,urlcolor=Blue]{hyperref}
\newcommand{\figref}[2]{\ref{#1}\hyperref[#1]{#2}}

\makeatletter
\def\NAT@spacechar{}
\makeatother
\begin{document}
\nocite{*}

\title{Sensitivity Enhancement in Atom-Interferometer Gyroscopes\\ via Phase-Modulation Signal Readout Scheme}

\author{Sotatsu~Otabe}
\email[]{sotatsu.otabe@phys.s.u-tokyo.ac.jp}
\altaffiliation[Present address: ]{Department of Physics, University of Tokyo, 7-3-1 Hongo, Bunkyo-ku, Tokyo 113-0033, Japan}
\affiliation{Institute of Integrated Research, Institute of Science Tokyo, 4259 Nagatsuta-cho, Midori-ku, Yokohama, Kanagawa 226-8501, Japan}

\author{Naoki~Kaku}
\affiliation{Department of Physics, Institute of Science Tokyo, 2-13-1 Ookayama, Meguro-ku, Tokyo 152-8550, Japan}

\author{Tomoya~Sato}
\affiliation{Institute of Integrated Research, Institute of Science Tokyo, 4259 Nagatsuta-cho, Midori-ku, Yokohama, Kanagawa 226-8501, Japan}

\author{Martin~Miranda}
\affiliation{Institute of Integrated Research, Institute of Science Tokyo, 4259 Nagatsuta-cho, Midori-ku, Yokohama, Kanagawa 226-8501, Japan}

\author{Takuya~Kawasaki}
\affiliation{Department of Physics, University of Tokyo, 7-3-1 Hongo, Bunkyo-ku, Tokyo 113-0033, Japan}

\author{Mikio~Kozuma}
\affiliation{Institute of Integrated Research, Institute of Science Tokyo, 4259 Nagatsuta-cho, Midori-ku, Yokohama, Kanagawa 226-8501, Japan}
\affiliation{Department of Physics, Institute of Science Tokyo, 2-13-1 Ookayama, Meguro-ku, Tokyo 152-8550, Japan}

\date{\today}

\begin{abstract}
Quantum sensors based on atom interferometers are advancing both fundamental physics and practical applications, with higher sensitivity being a key requirement for these investigations. Here, we experimentally demonstrate a sensitivity enhancement of an atom-interferometer gyroscope using a phase-modulation signal readout scheme. Phase modulation applied to the laser light used for atomic state manipulation is transferred to the atomic phase and read out via multi-harmonic demodulation. The observed sensitivity improvement factor of $1.20\pm0.04$ over the conventional phase sweep scheme agrees with theoretical predictions. We also found that phase-dispersion compensation control, which compensates atomic velocity dispersion and preserves interference contrast at high angular rates, effectively eliminates the nonlinearity inherent in multi-harmonic demodulation. The sensitivity improvement achieved by our method is applicable to a broad class of atom interferometers and requires no modifications to the optical or vacuum systems, making it particularly effective for size-constrained applications such as large-baseline experiments and inertial navigation systems.
\end{abstract}

\maketitle
\textit{Introduction.}---Light-pulse atom interferometers, which exploit the wave nature of matter for precise phase measurements, have found diverse applications ranging from precision sensing to fundamental physics, including measurements of rotation rate~\cite{PhysRevLett.67.177,PhysRevLett.78.2046}, gravity~\cite{PhysRevLett.67.181,Peters1999}, gravity gradients~\cite{PhysRevLett.81.971,PhysRevA.65.033608}, and fundamental constants~\cite{doi:10.1126/science.1135459,PhysRevLett.106.080801,Rosi2014,doi:10.1126/science.aap7706,Morel2020}. More recently, research has advanced these instruments toward increasingly ambitious frontiers~\cite{Buchmueller03042023}, including detection of gravitational waves~\cite{PhysRevLett.110.171102,Canuel2018,PhysRevD.101.124013,doi:10.1142/S0218271819400054}, dark matter~\cite{PhysRevD.105.023006,PhysRevD.106.095041}, or both~\cite{PhysRevD.97.075020,Badurina_2020,El-Neaj2020,Abe_2021,doi:10.1098/rsta.2021.0060,baynham2025prototypeatominterferometerdetect}, observation of the gravitational Aharonov-Bohm effect~\cite{doi:10.1126/science.abl7152}, tests of the weak equivalence principle~\cite{PhysRevLett.93.240404,PhysRevA.88.043615,PhysRevLett.112.203002,PhysRevLett.113.023005,PhysRevLett.115.013004,PhysRevLett.117.023001,Rosi2017,PhysRevLett.120.183604,PhysRevLett.125.191101,10.1116/5.0076502} and gravity theory~\cite{PhysRevLett.98.111102,PhysRevD.78.042003,Kovachy2015,PhysRevD.108.084038}, and searches for a fifth force~\cite{doi:10.1126/science.aaa8883,Jaffe2017,PhysRevLett.123.061102,Panda2024}. In parallel with these scientific developments, atom interferometers are increasingly demonstrated in realistic environments~\cite{Bongs2019,s23177651}, with mobile platforms~\cite{Schmidt2011,10.1063/1.4801756,Ménoret2018,Stray2022} and field-deployable systems~\cite{https://doi.org/10.1029/2022GL097814,PhysRevA.108.032811} being constructed or installed in vehicles, including cars~\cite{doi:10.1126/sciadv.aax0800,10.1063/5.0068761,s22166172}, ships~\cite{Bidel2018,Qiao2025}, aircraft~\cite{Geiger2011,Barrett2016,Bidel2020}, and spacecraft~\cite{Lachmann2021,Williams2024,10.1093/nsr/nwaf012}. In particular, the atom-interferometer gyroscope (AIG), which can produce a large Sagnac phase due to the slower longitudinal velocity and shorter wavelength of atoms compared to light, is emerging as a promising inertial sensor~\cite{PhysRevLett.67.177,PhysRevLett.78.2046,TLGustavson_2000,doi:10.1126/sciadv.abn8009}. Refinements in interferometer operation, enabled by advances in optical manipulation techniques~\cite{PhysRevLett.97.010402,PhysRevLett.97.240801,PhysRevLett.107.133001,PhysRevLett.116.183003,doi:10.1126/sciadv.aau7948,PhysRevApplied.23.044001,kraft2025phasemodulationdetectionstrontium}, have significantly improved AIG performance, accelerating its transition from fundamental research to practical sensors for inertial navigation.\par
The fundamental sensitivity of AIG is ultimately limited by atomic shot noise, i.e., the quantum projection noise in population detection used to read out the Sagnac phase, and sets the lower bound for angular random walk (ARW). Reducing ARW is essential for achieving navigation-grade gyroscope performance because lower ARW directly improves position and attitude accuracy~\cite{s17102284,El-Sheimy2020}. ARW can be suppressed by device improvements such as increasing atom flux, enhancing interferometer contrast, and expanding the effective area. Recently, we have theoretically proposed a phase-modulation signal readout scheme that can improve sensitivity compared to the conventional phase sweep scheme~\cite{PhysRevA.111.013302}. These readout schemes can be implemented by transferring the phase of Raman light, which is used to drive atomic transitions and generate superposition of momentum states, to atomic waves. In the phase sweep scheme, frequency detuning of the Raman light induces periodic variation in the interference fringes, from which the interferometric phase is extracted via orthogonal demodulation at the sweeping frequency. In contrast, the phase modulation scheme enables the readout of larger signals with reduced noise by applying phase modulation to the Raman light with an optimal modulation index near a suitable interference fringe. In a broad class of atom interferometers employing time-domain interference, an optical modulator adjusts the frequency and phase of the Raman light. Notably, the proposed phase modulation scheme can be implemented in such systems through software-based control of the Raman light using an optical modulator, requiring no significant hardware modifications.\par
In this Letter, we report the first experimental demonstration of sensitivity enhancement in an AIG using a phase modulation scheme. The experiment is conducted with an AIG employing thermal rubidium (Rb) atomic beams, and the phase modulation signal is applied to the optical modulator in the Raman light path. The interferometric phase is extracted through multi-harmonic demodulation and converted into a rotation rate. The angular velocity-equivalent amplitude spectral density (ASD) is evaluated to quantify the performance of AIG. Because the white noise floor in the ASD directly determines the ARW, we focus on this region to compare the phase modulation and phase sweep schemes.\par
\begin{figure}
	\centering
	\includegraphics[width=\hsize]{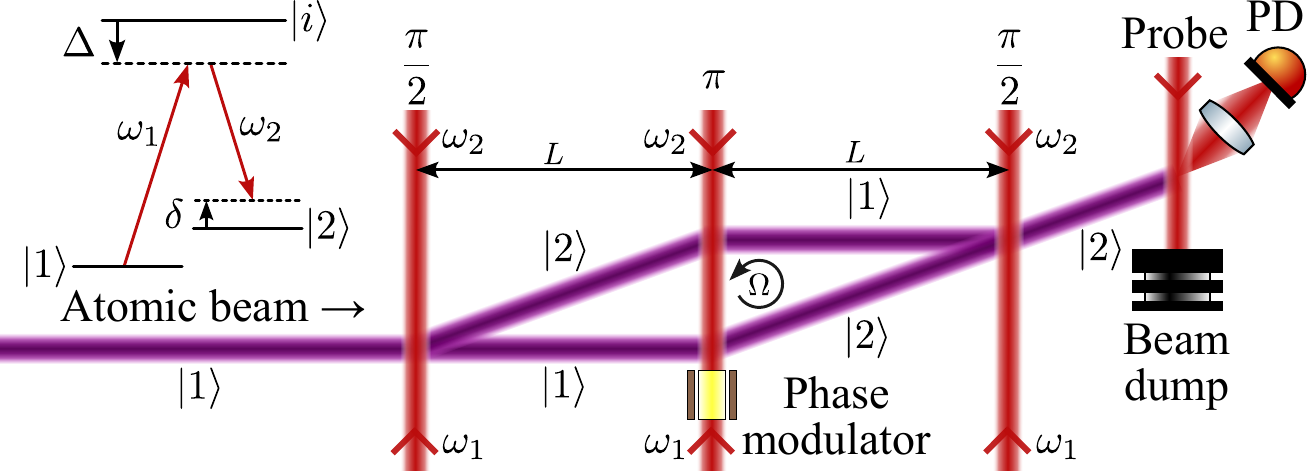}
	\caption{Schematic of atomic interferometry based on stimulated Raman transitions. Each Raman beam comprises upward- and downward-propagating components in the plane of the page, with frequencies $\omega_1$ and $\omega_2$, respectively. These frequencies are tuned to the three-level atomic system with the one-photon detuning $\Delta$ and two-photon detuning $\delta$.}
	\label{fig:outline}
\end{figure}
\textit{Rotation measurement with phase-modulated AIG.}---We investigate atomic interferometry using $\pi/2$--$\pi$--$\pi/2$ Raman pulses [Fig.\,\ref{fig:outline}]. The internal phase of the atomic Mach-Zehnder interferometer is differentially modulated by a phase modulator for the $\pi$ pulse. When the velocity distribution of atoms is neglected, the number of output atoms in energy level $\left|2\right>$, measured via fluorescence induced by the probe light and detected by a photodetector (PD), is proportional to
\begin{equation}
    P=\frac{1-\cos(\Phi)}{2},\label{eq:interferometerOutput}
\end{equation}
where $\Phi=\Phi_\mathrm{S}+\phi_\mathrm{m}$ is the total internal phase~\cite{berman1997atom,PMsupp}. Acceleration along the direction of Raman light propagation also induces phase shifts, which can be eliminated by constructing another interferometer using a counter-propagating atomic beam~\cite{TLGustavson_2000}. The Sagnac phase $\Phi_\mathrm{S}$, induced by system rotation, is given by
\begin{equation}
    \Phi_\mathrm{S}=\frac{2k_\mathrm{eff}\Omega L^2}{v},\label{eq:SagnacPhase}
\end{equation}
where $k_\mathrm{eff}$ is the magnitude of the effective two-photon wave vector for the Raman beams, $\Omega$ is the angular velocity of the system, $L$ is the pulse separation distance, and $v$ is the longitudinal velocity of atoms. The term $\phi_\mathrm{m}=2\beta\sin(\omega_\mathrm{m}t)+\omega_\mathrm{s}t+\phi_0$ represents the sum of phase modulation from the optical modulator and the phase sweep due to the two-photon detuning. Here, $\beta$ is the modulation index for the Raman beam, $\omega_\mathrm{m}/2\pi$ is the modulation frequency, $\omega_\mathrm{s}/2\pi$ is the sweep frequency, and $\phi_0$ is an additional phase that defines the operation fringe. When the phase sweep frequency is set to zero ($\omega_\mathrm{s}=0$), Eq.\,\eqref{eq:interferometerOutput} expands using Bessel functions of the first kind $J_n$:
\begin{align}
    P=& \frac{1}{2}\left(1-J_0(2\beta)\cos(\Phi_\mathrm{S}+\phi_0)\right) \nonumber\\
    +&\sum_{k=1}^\infty \left[J_{2k-1}(2\beta)\sin(\Phi_\mathrm{S}+\phi_0)\sin((2k-1)\omega_\mathrm{m}t)\right.\nonumber\\
    &\left.\ \ \ \ \ -J_{2k}(2\beta)\cos(\Phi_\mathrm{S}+\phi_0)\cos(2k\omega_\mathrm{m}t)\right].
\end{align}
The Sagnac phase can be read out by demodulating the signal at an integer multiple of $\omega_\mathrm{m}$.\par
To estimate the Sagnac phase with high tolerance to atomic number drift, the orthogonal signal must be read out, e.g., by demodulating at frequencies $\omega_\mathrm{m}$ and $2\omega_\mathrm{m}$. Experimentally, the amplitude of the signal read out from the atom interferometer depends on the demodulation frequency. One contributing factor is that the time required from modulation to demodulation depends on the longitudinal velocity of the atoms. If the atoms have velocity distribution, the optimal demodulation phase, at which the signal is maximized, varies across atoms, resulting in dephasing. The signals obtained by demodulating at frequencies $\omega_\mathrm{m}$ and $2\omega_\mathrm{m}$ are given by
\begin{align}
    X_1=&k_1(\omega_\mathrm{m})J_1(2\beta)\sin(\Phi_\mathrm{S}+\phi_0),\\
    X_2=&k_2(\omega_\mathrm{m})J_2(2\beta)\cos(\Phi_\mathrm{S}+\phi_0),
\end{align}
where $k_1(\omega_\mathrm{m})$ and $k_2(\omega_\mathrm{m})$ are dephasing factors for the first- and second-harmonic demodulated signals, respectively, with the demodulation phases adjusted to maximize each signal. When $\phi_0$ is controlled such that $P=0$ in the absence of Sagnac phase and modulation, i.e., when the interferometer operates at the dark fringe, the Sagnac phase can be estimated as
\begin{equation}
    \Phi_\mathrm{S}=\arctan\left(\frac{k_2(\omega_\mathrm{m})J_2(2\beta)}{k_1(\omega_\mathrm{m})J_1(2\beta)}\frac{X_1}{X_2}\right),\label{eq:SagnacPhaseEstimate}
\end{equation}
and then the angular velocity can be determined. This scheme preserves the zero-offset and high drift-resistance characteristics of the phase sweep scheme; however, errors in the dephasing factor directly introduce nonlinearity into the estimated angular velocity. As discussed later, this nonlinearity can be eliminated by phase-dispersion compensation control~\cite{PhysRevApplied.23.044001,Joyet2012}.\par
\begin{figure}
	\centering
	\includegraphics[width=\hsize]{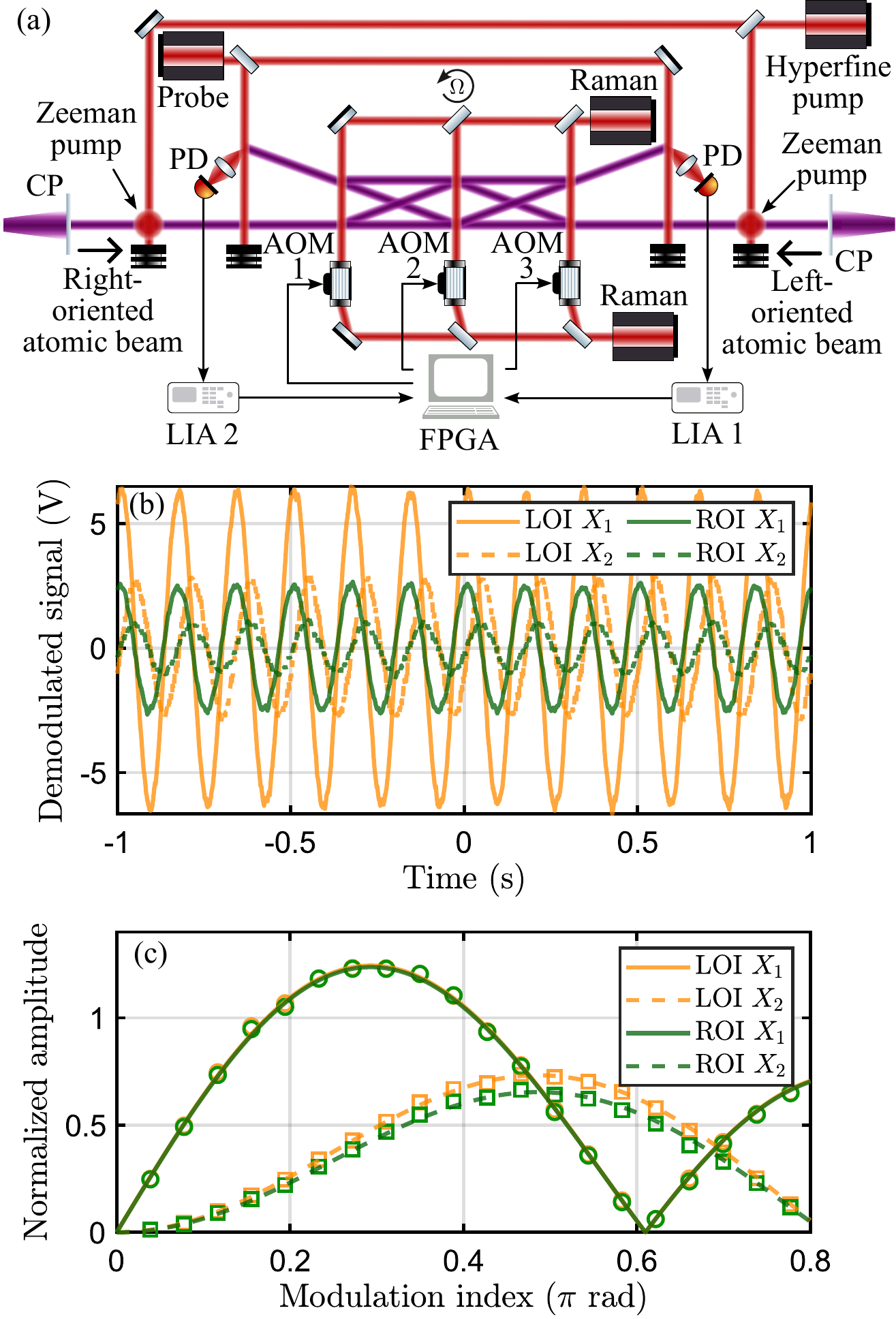}
	\caption{(a) Experimental setup. Atomic beams were optically pumped to the total angular momentum of $F=1$ and magnetic quantum number of $m_F=0$, then injected into the interference region comprising Raman beams 1, 2, and 3 with AOMs 1, 2, and 3. The phase of each upward-propagating Raman beam was controlled by detecting its beat signal with the corresponding downward-propagating beam. The probe light measured the number of outgoing atoms in $F=2$ via the $F=2$--$F'=3$ cyclic transition of $^{87}$Rb.
    (b) Demodulated signals obtained via the phase modulation with $\beta=0.3108 \pi$\,rad and phase sweep at $2\delta_2/2\pi=6$\,Hz.
    (c) Dependence of signal amplitude on modulation index. Circles (squares) and solid (dashed) lines represent the estimated signal amplitude of $X_1$ ($X_2$) obtained by the phase modulation scheme and the corresponding theoretical fit, respectively. The vertical axis is normalized by the signal amplitude obtained using the phase sweep and demodulation at $2\delta_2/2\pi=326$\,Hz and $320$\,Hz, respectively.
    }
	\label{fig:setup}
\end{figure}
\textit{Experimental setup.}---Figure\,\figref{fig:setup}{(a)} shows the experimental setup used to demonstrate the sensitivity enhancement in an atom interferometer employing the phase modulation scheme. The wavelength of all laser lights was approximately $780.2$\,nm, corresponding to the $D_2$ transition of $^{87}$Rb. We constructed left- and right-oriented interferometers (LOI and ROI) using bidirectionally propagating Rb atomic beams emitted from the crucibles heated to $105\pm2$\,$^\circ$C with temperature fluctuations within $0.3$\,$^\circ$C. Glass capillary plates (CPs) with a hole diameter of $5$\,\textmu m and thickness of $1$\,mm were used as collimators. By taking the phase difference between LOI and ROI, a phase component proportional to the angular velocity unaffected by acceleration was extracted~\cite{TLGustavson_2000,PhysRevApplied.23.044001}. The frequencies of the upward- and downward-propagating Raman beams were tuned to the $F=1$--$F'=0$ and $F=2$--$F'=0$ transitions of $^{87}$Rb, respectively, with a one-photon detuning of $\Delta/2\pi=1.5$\,GHz. The field-programmable gate array (FPGA) generates signals for the acousto-optic modulators (AOMs) based on the demodulated signals obtained from lock-in amplifiers (LIAs). We selected a modulation frequency of $\omega_\mathrm{m}/2\pi=320$\,Hz to achieve a fast data acquisition rate while maintaining low drift and high signal amplitude~\cite{PMsupp}.\par To confirm the demodulated signal, a small two-photon detuning was applied to Raman beam 2 ($2\delta_2/2\pi=6$\,Hz). In addition to the orthogonality of the measured signals $X_1$ and $X_2$, the phases of the LOI and ROI were closely matched by precisely adjusting the alignment of Raman beams [Fig.\,\figref{fig:setup}{(b)}]. Furthermore, to precisely determine the ratio of the dephasing factors $k_1(\omega_\mathrm{m})/k_2(\omega_\mathrm{m})$, we estimated the modulation index dependence of the signal amplitude by performing a fast Fourier transform on the demodulated signals from $10$\,s measurements and integrating over the range around approximately $5.5$--$6.5$\,Hz [Fig.\,\figref{fig:setup}{(c)}]. The estimated values of $k_1(\omega_\mathrm{m})/k_2(\omega_\mathrm{m})$ from the fitting were $1.420\pm0.006$ and $1.582\pm0.012$ for the LOI and ROI, respectively. By calibrating the ratio of $X_1$ to $X_2$ using these factors, as described in Eq.\,\eqref{eq:SagnacPhaseEstimate}, we can estimate the Sagnac phase in the atom interferometer while preserving linearity.\par
\begin{figure}
	\centering
	\includegraphics[width=\hsize]{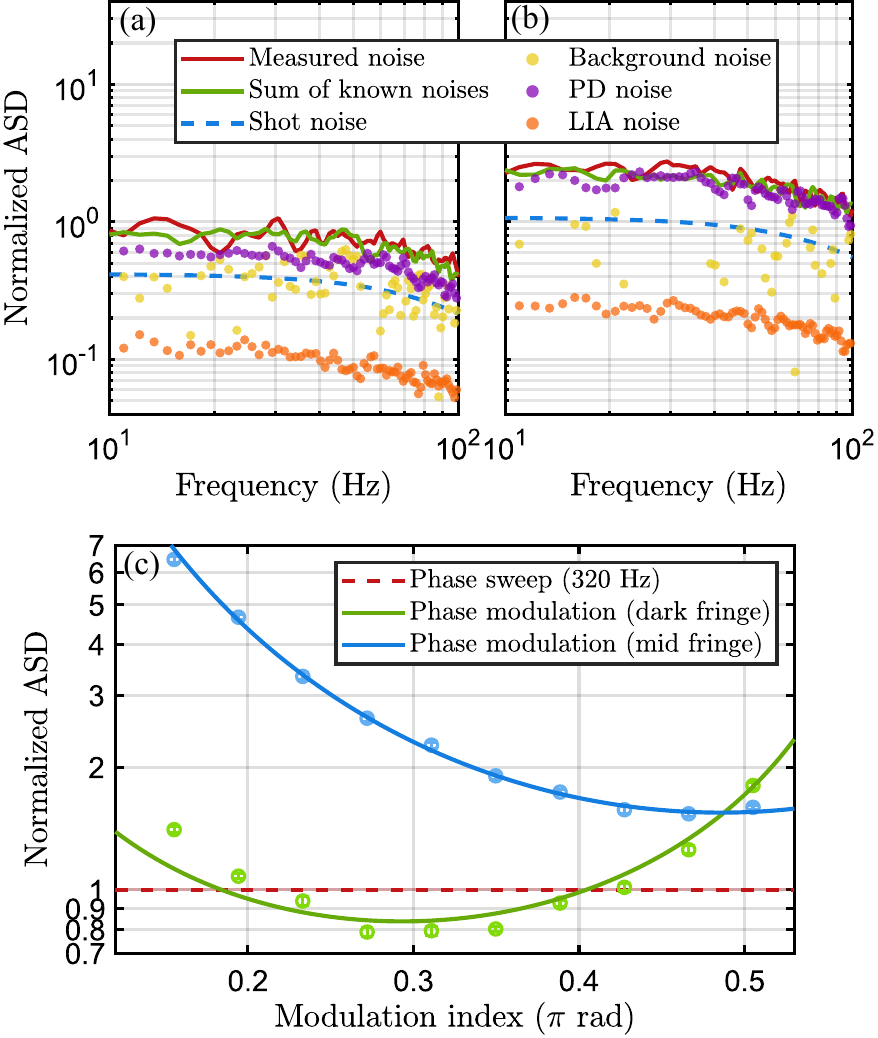}
	\caption{(a),(b) Noise budget of AIG for dark fringe and midfringe operation. Dotted lines indicate estimated shot noise, while circles represent measured noise levels. We set $\beta=0.3108\pi$\,rad to achieve near-optimal sensitivity at the dark fringe. ASD was calculated using the Welch method~\cite{1161901} from interferometer phase measurements acquired over $10$\,s at a sampling rate of $10$\,kHz. High-frequency signals were attenuated by a fourth-order low-pass filter with a cutoff frequency of $159$\,Hz in the LIA.
    (c) Sensitivity of the AIG using phase sweep and modulation schemes. Dotted lines and filled areas indicate the overall average of ASD and estimated standard error from $10$ measurements using a phase sweep scheme. Circles, error bars, and solid lines represent the average ASD in the $10$--$50$\,Hz range for $10$\,s measurements, estimated standard error, and theoretical fit for the phase modulation scheme. The vertical axis in (a), (b), and (c) is normalized by the average ASD obtained with the phase sweep scheme at $2\delta_2/2\pi=320$\,Hz.
    }
	\label{fig:results}
\end{figure}
\textit{Results.}---We investigated the angular velocity-equivalent ASD in the dark fringe [Fig.\,\figref{fig:results}{(a)}] and midfringe [Fig.\,\figref{fig:results}{(b)}] to identify the noise sources limiting AIG sensitivity. In both cases, the measured noise agreed well with the sum of known noises. Shot noise was estimated by measuring the photodiode current with Raman beams 1 and 3 blocked, thereby maximizing the number of $F=2$ atoms without interference~\cite{PMsupp}. Background fluorescence noise was measured by turning off all Raman laser lights. This noise arises from fluctuations in the number of atoms not undergoing Raman transitions, as confirmed by noise injection into the pump light power~\cite{PMsupp}. PD and LIA noise were obtained by turning off all laser lights and disconnecting signals to the LIAs, respectively. In summary, PD, background fluorescence, and shot noise were dominant at the dark fringe operation, while only PD noise was dominant at the midfringe operation. Background and PD noise originated from down-converted fluctuations near the demodulation frequency and were independent of the modulation index~\cite{PMsupp}. Although shot noise could depend on the modulation index~\cite{PhysRevA.111.013302}, this effect is negligible due to the low interferometer contrast, which was approximately $2\%$ for both ROI and LOI~\cite{PMsupp}. The AIG sensitivity was evaluated by averaging the ASD over the $10$--$50$\,Hz range, where PD, background, and shot noise can be treated as white noise. \par
Figure\,\figref{fig:results}{(c)} compares AIG sensitivity obtained using the phase sweep and modulation schemes. Because the noise affecting AIG sensitivity may drift during measurements, the ASD for the phase sweep scheme was measured over $1000$\,s. This time is comparable to the duration required to measure modulation index dependence of sensitivity in the phase modulation scheme. We extracted $10$\,s of data every $100$\,s and calculated the average ASD over $10$--$50$\,Hz for each segment, confirming that temporal sensitivity variations were negligible. The measured sensitivity using the phase modulation scheme at the dark fringe was clearly improved relative to the phase sweep scheme with $2\delta_2/2\pi=320$\,Hz for modulation index over approximately $0.2$--$0.4\pi$\,rad. Assuming dominant noise sources were independent of the modulation index, optimal sensitivity was obtained at $\beta\simeq0.29\pi$\,rad, with an estimated sensitivity improvement factor of $1.20\pm0.04$ ($0.80\pm0.14$\,dB). In contrast, optimal sensitivity at midfringe, obtained at $\beta\simeq0.49\pi$\,rad, was significantly worse than that of the phase sweep scheme by a factor of $0.646\pm0.008$ ($-1.89\pm0.05$\,dB). The deviation from the theoretical curve observed at low modulation indices in the dark fringe was attributed to fringe setting errors~\cite{PMsupp}.\par
\begin{figure}
	\centering
	\includegraphics[width=\hsize]{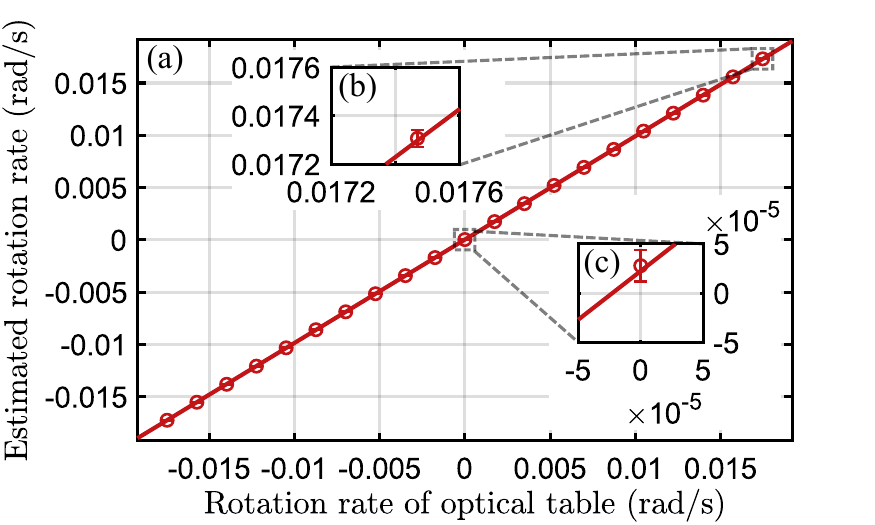}
	\caption{(a) Rotation rate estimation using the phase modulation scheme. The horizontal and vertical axis values were measured by BlueSeis3A, a fiber-optic gyroscope (FOG) with a scale factor variation of approximately $50$\,ppm~\cite{10.1785/0220170143}, and our AIG, respectively, with an averaging time of $3$\,s. Each FOG measurement was offset by the non-rotating measurement to eliminate the angular velocity due to Earth's rotation. The error bars on the vertical axis represent the standard error for $10$ measurements. (b),(c) Expanded plot around $0.0174$ and $0$\,rad/s.}
	\label{fig:angularVelocity}
\end{figure}
When the angular velocity is determined directly from the estimated phase using multi-harmonic demodulation, errors in the dephasing factor introduce nonlinearity into the measured angular velocity. By rotating the optical table, we demonstrated that phase-dispersion compensation control, which suppresses rotation-induced dephasing~\cite{PhysRevApplied.23.044001,Joyet2012}, effectively eliminates this nonlinearity. In addition to activating Mach–Zehnder interferometer phase control, we implemented phase-dispersion compensation control on the frequencies of Raman beams 1 and 3, with an effective cutoff frequency of $22.1\pm0.1$\,Hz~\cite{PMsupp}. The measured angular velocity exhibits linearity, with low measurement error and residuals from the fit [Figs.\,\figref{fig:angularVelocity}{(a)} and \figref{fig:angularVelocity}{(b)}]. The estimated phase is used solely for contrast restoration through feedback, without affecting angular velocity estimation, thereby preserving robustness against phase nonlinearities introduced by the modulation scheme. The estimated slope and intercept from the linear fit are $0.9890\pm0.0002$ and $(2.2\pm0.3)\times10^{-5}$\,rad/s, respectively. The deviation of the slope from unity arises from a mismatch between the actual and design pulse separations. This fitting process corresponds to the calibration of AIG, yielding an estimated pulse separation distance of $(6.923\pm0.002) \times 10^{-2}$\,m. Using this calibration, the angular velocity-equivalent ASD averaging over $1$--$10$\,Hz was determined to be $(7.7\pm0.1)\times10^{-6}$\,rad/s/$\sqrt{\mathrm{Hz}}$, based on $10$ measurements with a duration time of $100$\,s. A small but non-zero intercept [Fig.\,\figref{fig:angularVelocity}{(c)}] may result from the target phase estimation process in phase-dispersion compensation control~\cite{PMsupp}.\par
\textit{Discussion.}---When the sensitivity of atomic interferometry is limited by modulation index-independent noise and velocity distributions are negligible, the sensitivity ratio between phase sweep and modulation schemes is given by $2J_1(2\beta)$ for dark fringe operation~\cite{PhysRevA.111.013302,PMsupp}, with a maximum value of approximately $1.16$. The experimentally observed sensitivity improvement is consistent with this theoretical value within the margin of error. Differences in signal amplitude and dephasing factor between LOI and ROI arose from misalignments in the angle settings of the atomic source and pump beams, which affected the number and velocity distribution of atoms reaching the interferometer region. Under these conditions, it is analytically challenging to identify a modulation frequency that simultaneously yields high signal amplitude and suppresses drift. Therefore, we plan to develop a simulator that models the behavior of individual atoms.\par
The phase modulation scheme, capable of large-signal readout at appropriate modulation indices, can achieve better sensitivity than the phase sweep scheme under various fundamental and technical noise sources. In particular, the phase modulation scheme is expected to significantly enhance the sensitivity of shot-noise-limited atom interferometers~\cite{PhysRevA.80.063604,baynham2025prototypeatominterferometerdetect,PhysRevA.65.033608}, with a sensitivity improvement ratio of up to approximately $1.63$~\cite{PhysRevA.111.013302}. To further improve the sensitivity with the phase modulation scheme, a technique using a modulation and demodulation with third or higher harmonics has been applied to optical interferometry~\cite{PhysRevA.43.5022, Gray:93, s23094442}; however, this technique may yield limited benefits in atomic interferometry due to strong dephasing at higher demodulation frequencies.\par
Our method offers practical advantages, as it can be implemented straightforwardly by modifying the drive signal of the optical modulator commonly installed in atom interferometers. Notably, phase modulation using an AOM has recently been implemented in AIG employing a thermal strontium atomic beam~\cite{kraft2025phasemodulationdetectionstrontium}. In contrast, the Kerr modulator for the atom interferometer~\cite{PhysRevLett.116.053004,Décamps2017} requires redesigning the optical and vacuum systems. Although the multi-harmonic demodulation required for the phase modulation scheme introduces nonlinearity in Sagnac phase estimation, it can be calibrated using the dephasing-factor estimation method. Additionally, phase-dispersion compensation control is a key technique for phase-modulated AIG because it suppresses rotation-induced dephasing and eliminates the impact of phase nonlinearity on angular velocity estimation. The developed control system, which enables rotational measurements while maintaining linearity, is detailed in the Supplemental Material~\cite{PMsupp}.\par
\textit{Conclusion.}---We experimentally demonstrate sensitivity improvement in an AIG using a phase-modulation signal readout scheme. Our approach achieves a rotation rate sensitivity of $7.7\pm0.1$\,\textmu rad/s/$\sqrt{\mathrm{Hz}}$ in the $1$--$10$\,Hz range, accompanying a sensitivity enhancement of $1.20\pm0.04$ over the conventional phase sweep scheme. Sensitivity improvement from the phase modulation scheme would be more pronounced in the shot-noise-limited interferometers. We believe this approach is particularly useful in scenarios requiring high measurement accuracy under size constraints, such as inertial navigation systems and large-baseline setups for fundamental physics research.\par
\textit{Acknowledgments.}---This work was supported by the Japan Science and Technology Agency (JST) under Grants JPMJMI17A3 and JPMJPF2015.\par

\bibliography{PMbib}

\end{document}